\documentclass[runningheads]{llncs}
\usepackage{graphicx}
\usepackage{comment}
\usepackage{cite}
\usepackage{mathtools}
\usepackage{amsmath,amssymb,amsfonts}
\usepackage{graphicx}
\usepackage{textcomp}
\usepackage{xcolor}
\usepackage{amsmath}
\usepackage{multirow}
\usepackage{multicol}
\usepackage{caption}
\usepackage{array}
\usepackage{subcaption}
\usepackage{balance}
\usepackage{wrapfig}
\usepackage{hyperref}
\usepackage{multirow}
\usepackage{multicol}
\usepackage{color}
\usepackage{parskip}
\usepackage{colortbl}
\usepackage{booktabs}
\usepackage{flushend}
\usepackage{todonotes}
\usepackage{cite}
\usepackage{siunitx}
\usepackage{ulem} 

\usepackage[utf8]{inputenc}
\usepackage[english]{babel}

\usepackage{algpseudocode}
\usepackage{algorithm}

\newcolumntype{L}{>{\centering\arraybackslash}m{3cm}}

\makeatletter
\newcommand{\linebreakand}{%
  \end{@IEEEauthorhalign}
  \hfill\mbox{}\par
  \mbox{}\hfill\begin{@IEEEauthorhalign}
}
\makeatother

\def\BibTeX{{\rm B\kern-.05em{\sc i\kern-.025em b}\kern-.08em
    T\kern-.1667em\lower.7ex\hbox{E}\kern-.125emX}}
\begin{document}
\title{Pain Intensity Assessment in Sickle Cell Disease patients using Vital Signs during Hospital Visits\thanks{This work was supported by the NIH under grant NIH R01AT010413 }}
\titlerunning{SCD Pain intensity assessment using vital signs during hospital visits}

\author{Swati Padhee\inst{1}\and
Amanuel Alambo\inst{1}\and
Tanvi Banerjee \inst{1} \and
Arvind Subramaniam \inst{2} \and
Daniel M.~Abrams\inst{3} \and 
Gary K.~Nave, Jr. \inst{3} \and
Nirmish Shah \inst{2}}
\authorrunning{Padhee et al.}

\institute{Wright State University\and
Duke University \and
Northwestern University}
\maketitle         

\begin{abstract}

Pain in sickle cell disease (SCD) is often associated with increased morbidity, mortality, and high healthcare costs. The standard method for predicting the absence, presence, and intensity of pain has long been self-report. However, medical providers struggle to manage patients based on subjective pain reports correctly and pain medications often lead to further difficulties in patient communication as they may cause sedation and sleepiness. Recent studies have shown that objective physiological measures can predict subjective self-reported pain scores for inpatient visits using machine learning (ML) techniques. In this study, we evaluate the generalizability of ML techniques to data collected from 50 patients over an extended period across three types of hospital visits (i.e., inpatient, outpatient and outpatient evaluation). We compare five classification algorithms for various pain intensity levels at both intra-individual (within each patient) and inter-individual (between patients) level. While all the tested classifiers perform much better than chance, a Decision Tree (DT) model performs best at predicting pain on an 11-point severity scale (from 0-10) with an accuracy of 0.728 at an inter-individual level and 0.653 at an intra-individual level. The accuracy of DT significantly improves to 0.941 on a 2-point rating scale (i.e., no/mild pain: 0–5, severe pain: 6–10) at an inter-individual level. Our experimental results demonstrate that ML techniques can provide an objective and quantitative evaluation of pain intensity levels for all three types of hospital visits.

\end{abstract}


\section{Introduction}
\label{intro}
Sickle cell disease (SCD) is the most common inherited blood disorder, affecting millions of people worldwide. It is characterized by the production of an altered type of hemoglobin. The altered hemoglobin deoxygenates while passing through blood vessels, polymerizes and becomes fibrous, causing the red blood cells to become rigid and change their shape to sickle-shaped. The altered red blood cells can occlude blood vessels, a phenomenon known as vaso-occlusion, resulting in a lack of oxygen to tissues, and thereby causing pain \cite{manwani2013vaso}. Most patients with SCD experience repeated, unpredictable episodes of severe pain. These pain episodes are the leading cause of emergency department visits and may last for as long as several weeks. Arguably, the most challenging aspect of treating pain episodes in SCD is assessing and interpreting the patient’s pain intensity level.

However, in current clinical practice, self-description is the gold standard approach for determining the absence, presence, and intensity of pain.
Due to the subjective nature of pain, it becomes challenging for the clinicians to precisely ascertain the severity of the patient’s pain. Besides, effective treatment is palliative, including intravenous opioid therapy. While these self-described pain intensity levels provide important clinical reference indicators and have been proven to be useful for treating patients suffering from pain in most situations \cite{brown2011towards}, it might have challenges when applied to certain vulnerable populations. 

Current clinical guidelines recommend frequent observations of vital signs during assessment and treatment of painful episodes as they are an objective measurement for the essential physiological functions and are potential indicators for patients’ subjective pain levels. 
It has been previously reported that ML techniques can be used to design objective pain assessment models using vital signs from inpatient EHR data. However, people with SCD suffer from various acute complications that can result in multiple hospitalizations, emergency department (ED) visits, and outpatient care visits. To the best of our knowledge, this is the first study to explore the relationship between the varying nature of hospital visits and physiological measures on pain intensity for patients with SCD.


\section{Related Work}
\label{related}
Current literature shows increased attention on machine learning techniques to understand various complexities associated with patient health in SCD. 
Milton et al. \cite{milton2014prediction} developed an ensemble model exploring 14 algorithms to predict Hemoglobin F (HbF) in patients associated with different configurations of Single Nucleotide Polymorphisms (SNPs). 
Allayous et al. \cite{allayous2008machine} demonstrated the high risk of an acute splenic sequestration crisis, which is a severe symptom of SCD. Solanki \cite{solanki2014data}, implemented two models, including DT to classify specific blood groups. Khalaf et. al. \cite{khalaf2017machine} classified the dosage of medication required for the treatment of patients with SCD. 

Prior studies have reported that fluctuations in vital signs can be used for assessing pain \cite{arbour2014can} as acute pain leads to changes in vital signs \cite{macintyre2010acute}. These physiological measures include blood pressure, respiratory rate, oxygen saturation, temperature, and pulse. 
From our research group, Yang et.al. \cite{yang2018improving} showed the feasibility of ML techniques on a limited dataset of 5363 records from 40 patients during inpatient hospital visits to predict subjective pain scores from six objective vital signs. 
Alambo et. al. \cite{alambo2020measuring} employed 424 clinical notes of the same cohort of 40 patients to assess the prevalence of pain in patients and whether pain increases, decreases, or stays constant. In this study, we investigate the generalizability of ML techniques for a broader group of people during inpatient, outpatient, and outpatient evaluation visits. We provide definitions of these visits as validated by our clinical collaborators in Section definitions \ref{visit_definition}. Specifically, we utilize five years of EHR data from 50 patients suffering from SCD to build pain prediction models using objective physiological measures as features at both the intra-individual and inter-individual levels based on an 11-point numeric rating scale (NRS) \cite{downie1978studies}. We further investigate whether the variation in the type of hospital visits affect our model performance.


\section{Methods}
\label{mm}
\subsection{Data Description}
\label{data}
In this study, we utilized 67927 records from EHR data collected from 50 participants at Duke University Hospital over five consecutive years. 
Each record contained measures for six vital signs as follows: (i) peripheral capillary oxygen saturation (SpO2), (ii) systolic blood pressure (SystolicBP), (iii) diastolic blood pressure (DiastolicBP), (iv) heart rate (Pulse), (v) respiratory rate (Resp), and (vi) temperature (Temp). Along with the vital signs, each record also included the patient's self-reported pain score with an ordinal range from 0 (no pain) to 10 (severe and unbearable pain). 

The data were de-identified using study labels to label the patient without identification. The timestamp for each data entry was also de-identified, preserving temporality. 
The dataset had missing values for one or more of the vital signs and the pain score. Our analysis is done on 59728 records containing at least one of the six vital signs or pain score values from 47 patients as we observed that no data was extracted for three patients. 
As the percentage of complete records in our dataset was only 7.6\%, we employed an imputation method to impute the missing data values. We utilized Multiple Imputations by Chained Equations \cite{azur2011multiple}, sometimes called ``fully conditional specification" or ``sequential regression multiple imputations," as it is widely used in clinical practice and recent healthcare studies \cite{shah2015type,yang2018improving}.
\begin{figure}[!tbp]
  \centering
  \begin{minipage}[b]{0.58\textwidth}
    \includegraphics[width=\textwidth]{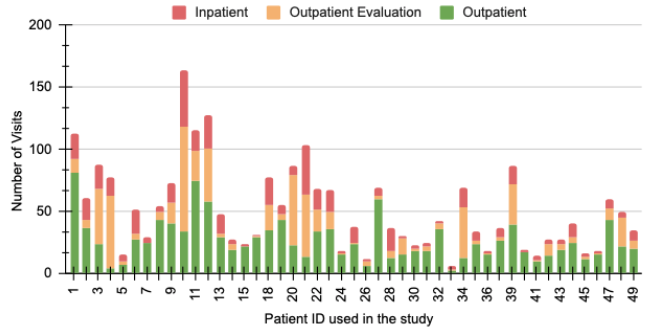}
    \vspace{-2.5em}
    \caption{Distribution of visits for 50 patients. (Study Patient identifiers 17, 37 and 50 are absent.)}
    \label{visits}
  \end{minipage}
  \hfill
  \begin{minipage}[b]{0.4\textwidth}
    \includegraphics[width=\textwidth]{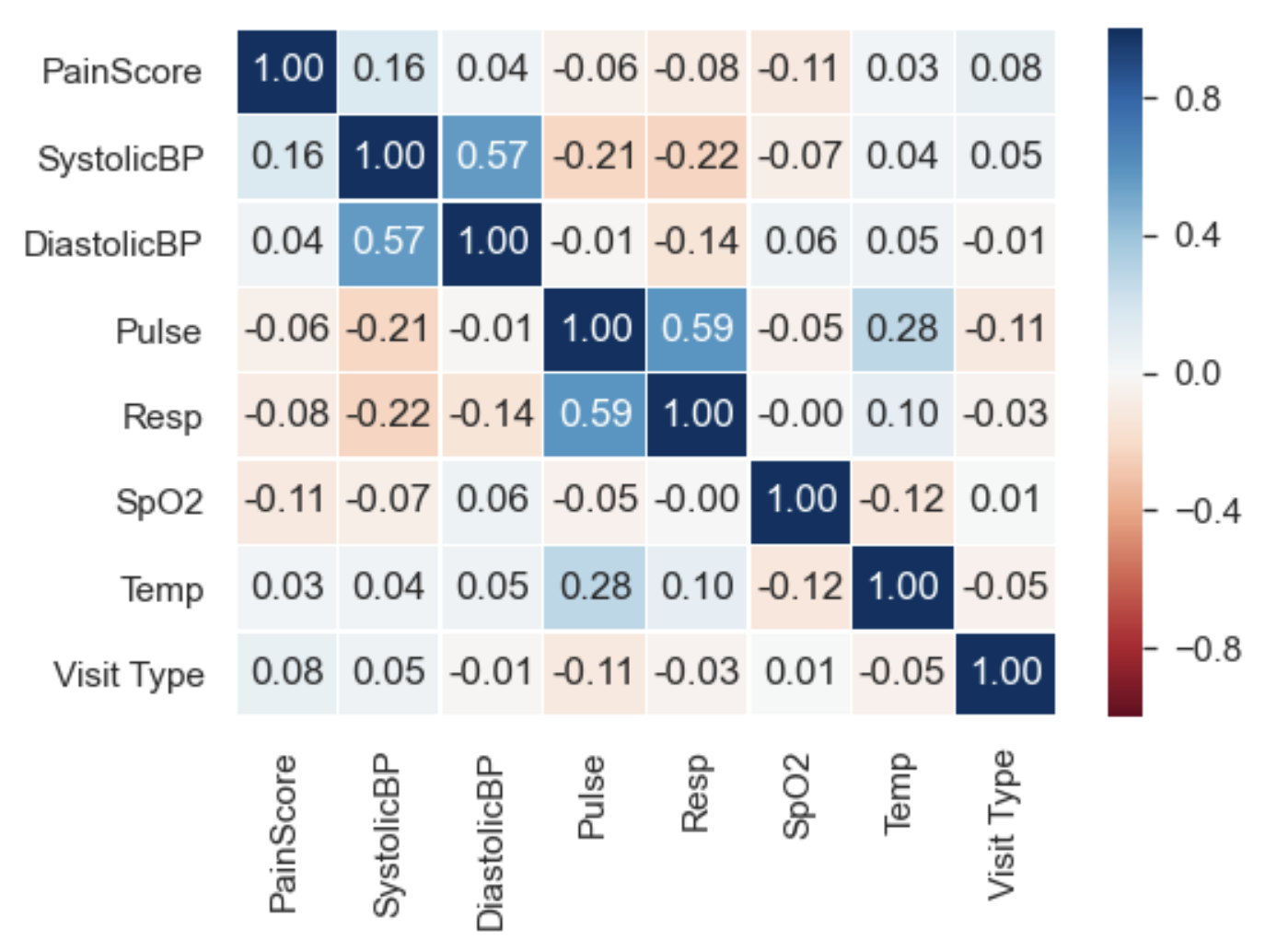}
    \vspace{-2.5em}
    \caption{Pearson correlation of six vital signs and visit types in the original dataset. }
    \label{pearson}
  \end{minipage}
\end{figure}

\subsubsection{Type of Hospital Visit:}
\label{visit_definition}
Some patients with SCD have higher inpatient requirements than others due to the subjectivity and frequency of pain crisis. Furthermore, because of SCD-related complications, many people with SCD may visit hospitals more frequently. 
However, limited information is available related to various hospital visits' characteristics, including emergency department visits among SCD patients. Information related to the type of hospital visit by patients with SCD can help develop services and strategies for best meeting patients' healthcare needs with SCD. To understand the variations in the nature of visits, we followed the definitions below recommended by our co-author clinician to extract information about the nature of visits for every record in our dataset.
\begin{itemize}
  \item \textbf{Visit}: For each patient, we consider a record to be of a different visit if there is a gap of at least two days between the records.
  \item \textbf{Outpatient visit}: We define a visit to be an outpatient visit if the patient has not stayed in the hospital for a day or longer and has two or less recordings taken.
  \item \textbf{Inpatient visit}: We define a visit to be an inpatient visit if the patient has stayed for two or more consecutive days in the hospital.
  \item \textbf{Outpatient evaluation visit}: We define a visit to be an outpatient evaluation visit if the patient has stayed in the hospital for one day or has more than two recordings taken in a single day.
\end{itemize}

Figure \ref{visits} shows the distribution of the three types of visits in our data.



\subsection{Pain Prediction}
\label{pc}
We examined the Pearson correlation between the six vital signs and the type of visit in our dataset as we plan to use them as features influencing pain scores. As shown in Figure \ref{pearson}, in addition to a moderate correlation of 0.57 between systolic and diastolic blood pressure, we observe a correlation of 0.59 between pulse and respiratory rate in the original dataset. The other variables are poorly correlated or uncorrelated with one another. Hence, we utilize all six vital signs and visit information as predictors of our pain prediction models.  
\begin{figure}[!tbp]
  \centering
  \begin{minipage}[b]{0.45\textwidth}
    \includegraphics[width=\textwidth]{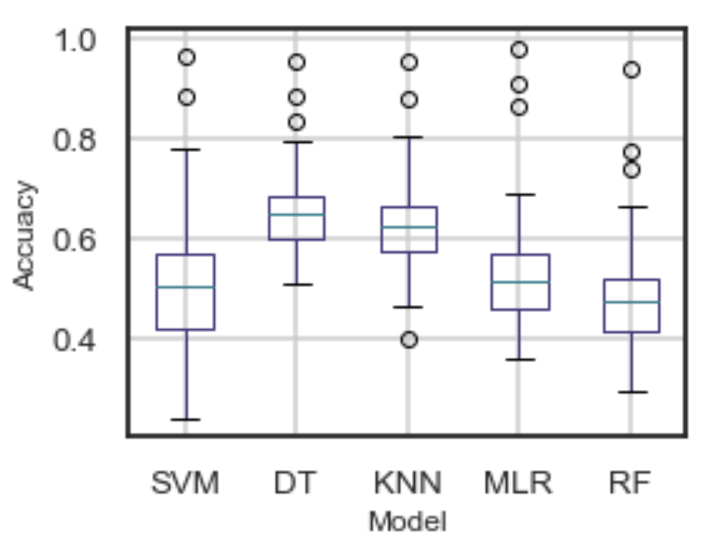}
    \vspace{-2.5em}
    \caption{Intra-individual pain prediction accuracy results on vital signs data.}
    \label{intra_v_plot}
  \end{minipage}
  \hfill
  \begin{minipage}[b]{0.45\textwidth}
    \includegraphics[width=\textwidth]{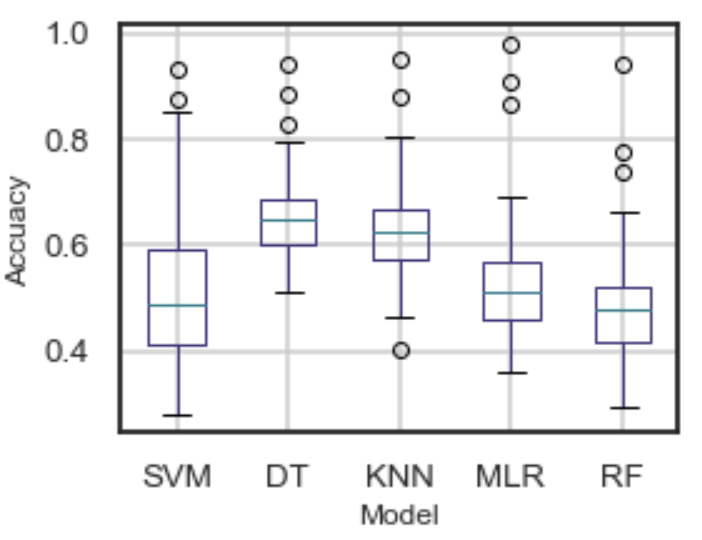}
    \vspace{-2.5em}
    \caption{Intra-individual pain prediction accuracy results on vital signs and visit information.}
    \label{intra_vv_plot}
  \end{minipage}

\end{figure}

\begin{table*}
\centering
\vspace{-2.0em}
\caption{Intra-individual pain prediction results (accuracy)}
\label{intra}
\scalebox{1.0}{
\begin{tabular}{|l|lllll|}
\cline{1-6}
               & SVM   & DT    & kNN   & MLR   & RF    \\
\cline{1-6}
Vitals         & 0.522	& \textbf{0.653}	&0.625	&0.535	&0.485\\
Vitals + Visit & 0.506	& \textbf{0.653}	& 0.625	&0.535	&0.486\\
Yang e.t al. \cite{yang2018improving}     & 0.582 & \hfill{\textemdash}   & 0.522 & 0.578 & 0.523\\
\cline{1-6}
\end{tabular}%
}
\end{table*}

We implemented five supervised ML classification algorithms to predict patients’ pain scores based on their vital signs: k-Nearest Neighbors (kNN), Support Vector Machine (SVM), Multinomial Logistic Regression (MLR), Decision Tree (DT), and Random Forest (RF). We investigated both the intra-individual level and inter-individual level (i.e., treating all the patients as a single entity) analysis. For intra-level analysis, we used the six vital signs, visit information, and pain scores as patient labels were unused since samples from the same patient were employed to build the personal model in this analysis. However, for the inter-individual level analysis, we employed the four different scenarios as reported by Yang et al. \cite{yang2018improving} i.e., Case 1: imputation with patient labels and prediction with patient labels; Case 2: imputation with patient labels and prediction without patient labels; Case 3: imputation without patient labels and prediction with patient labels; Case 4: imputation without patient labels and prediction without patient labels. For each experiment, we report the results with and without visit information. We used 10-fold cross-validation to evaluate our prediction models. We reported the model prediction accuracy as it is the ratio of correctly predicted pain scores over the total number of pain scores.

\begin{table}[h]

\centering
\caption{Inter-individual pain prediction results (accuracy)}
\label{inter}
\scalebox{0.9}{
\vline
\begin{tabular}{c|ccccc|ccccc|cc|}
\cline{1-13}
\multicolumn{1}{c}{}
\vline
&\multicolumn{5}{c}{Vitals}
\vline
& \multicolumn{5}{c}{Vitals + Visit}
\vline
& \multicolumn{2}{c}{Yang et. al.\cite{yang2018improving}}
\vline
\\
\multicolumn{1}{c}{}
\vline
& \multicolumn{1}{c}{SVM}
& \multicolumn{1}{c}{DT}
& \multicolumn{1}{c}{kNN}

& \multicolumn{1}{c}{MLR}
& \multicolumn{1}{c}{RF}
\vline
& \multicolumn{1}{c}{SVM}
& \multicolumn{1}{c}{DT}
& \multicolumn{1}{c}{kNN}

& \multicolumn{1}{c}{MLR}
& \multicolumn{1}{c}{RF}
\vline
& \multicolumn{1}{c}{MLR}
& \multicolumn{1}{c}{SVM}
\vline
\\ \cline{1-13}
Case 1               & 0.585 & 0.676  & 0.662          & 0.448 & 0.336       &0.595 & \textbf{0.697} &0.668  &0.525 &0.357     &0.429 &0.421\\
Case 2               & 0.422 & \textbf{0.647} & 0.644         & 0.345 & 0.335       &0.593 &0.643  &0.530 &0.427 &0.335     &0.215 &0.236\\
Case 3               & 0.561 & \textbf{0.701} & 0.658          & 0.405 & 0.406       &0.591 &0.701  &0.595 &0.460 &0.410     &0.313 &0.305\\
Case 4               & 0.590  & \textbf{0.728} & 0.708 & 0.401 & 0.404       &0.659 &0.704  &0.595 &0.472 &0.401     &0.257 &0.246\\
\cline{1-13}
\end{tabular}
}

\centering
\caption{Inter-individual pain prediction results with varying pain scales on vitals data (accuracy) arranged from higher resolution to lower resolution.
(6 Pain Scores: None:0, Very mild:1-2,Mild: 3–4, Moderate: 5–6, Severe: 7–8, Very severe:9-10;
4 Pain Scores: None: 0,Mild: 1–3,Moderate: 4–6 ,Severe: 7–10;
2 Pain Scores:No/mild Pain: 0–5, Severe Pain: 6–10)}
\label{inter_scale}
\scalebox{0.8}{
\begin{tabular}{|c|cccccc|cccccc|}
\hline
\multicolumn{1}{|l|}{} & \multicolumn{6}{c|}{11 Pain Score}              & \multicolumn{6}{c|}{6 Pain Score}                        \\ 
\multicolumn{1}{|l|}{} & SVM   & DT    & kNN            & MLR   & RF    &Yang et.al.\cite{yang2018improving} & SVM           & DT    & kNN            & MLR    & RF    &Yang et.al.\cite{yang2018improving}\\ \hline
Case 1               & 0.585 & 0.676  & 0.662          & 0.448 & 0.336   &0.429                               & 0.767         & 0.779  & 0.761         & 0.606  & 0.481 &0.546\\
Case 2               & 0.422 & 0.647 & 0.644          & 0.345 & 0.335   &0.215                               & 0.672         & 0.762 & 0.732          & 0.55   & 0.485 &0.347\\
Case 3               & 0.561 & 0.701 & 0.658          & 0.405 & 0.406   &0.313                               & 0.766         & 0.771 & 0.773         & 0.599  & 0.586 &0.449\\ 
Case 4               & 0.590  & \textbf{0.728} & 0.708 & 0.401 & 0.404   &0.257                               & 0.772         & \textbf{0.814} &0.777  & 0.605  & 0.589 &0.397\\ \hline
\multicolumn{1}{|l|}{} & \multicolumn{6}{c|}{4 Pain Score}               & \multicolumn{6}{c|}{2 Pain Score}                        \\ 
\multicolumn{1}{|l|}{} & SVM   & DT    & kNN          & MLR   & RF      &Yang et.al.\cite{yang2018improving} & SVM           & DT    & kNN            & MLR    & RF         &Yang et.al.\cite{yang2018improving}\\ \hline
Case 1               & 0.849 & 0.832  & 0.809         & 0.683 & 0.583   &0.681                            & 0.923        & 0.937  & 0.904         & 0.926  & 0.84           &0.821        \\
Case 2               & 0.788 & 0.821   & 0.788        & 0.659 & 0.589   &0.521                            & 0.915         & 0.919  & 0.893         & 0.903  & 0.835         &0.680          \\
Case 3               & 0.837 & 0.824 & 0.815          & 0.685 & 0.66    &0.607                            & 0.923         &0.939   & 0.907        & 0.9267 & 0.874          &0.730              \\
Case 4               & 0.85  & \textbf{0.853} &0.818  & 0.687 & 0.671   &0.563                            & 0.935         & \textbf{0.941} & 0.907  & 0.927  & 0.871        &0.678          \\ \hline
\end{tabular}%
}

\centering
\caption{Pain change prediction results (accuracy)}
\label{pain_change}
\scalebox{1.0}{
\begin{tabular}{|l|lll|lll|l|}
\hline
      & Vitals         &       &       & \multicolumn{3}{l|}{Vitals + Visit} & Yang. et. al \cite{yang2018improving} \\ 
      & DT             & kNN   & MLR   & DT               & kNN    & MLR    & MLR                                   \\ \hline
Case1 & 0.515          & 0.490 & 0.514 & \textbf{0.522}   & 0.504  & 0.517  & 0.403                                 \\
Case2 & \textbf{0.508} & 0.494 & 0.508 & 0.518            & 0.494  & 0.503  & 0.363                                 \\
Case3 & \textbf{0.518} & 0.466 & 0.517 & 0.520            & 0.492  & 0.518  & 0.390                                 \\
Case4 & \textbf{0.520} & 0.492 & 0.520 & 0.517            & 0.466  & 0.516  & 0.404                                \\ \hline
\end{tabular}%
}

\end{table}

\section{Results And Discussion}
\label{results}

\subsection{Intra-individual Pain Prediction}
\label{intra-result}

We present the intra-individual pain prediction results for 47 patients in terms of accuracy in Table \ref{intra}. Figure \ref{intra_v_plot} shows the accuracy distribution of predictions for all five classifiers. DT achieved the highest accuracy ranging from 0.503 to 0.953, and an average accuracy of 0.653 when trained on the six vital signs described in Section \ref{data}. Thus, our models trained on the vital signs of a patient could correctly predict the self-reported pain scores of the same patient on an average of 65.3\% of the records.  
Similar performance with additional visit information (Table \ref{intra}, row 2) indicates that for the same patient, our models can learn the differences between vital signs and pain intensity experienced by a patient during different types of visits. Our results show that a model trained on the same patient's historical data can predict the pain intensity levels for the same patient in the future based on their vital signs during outpatient, inpatient, and outpatient evaluation visits. Such a model can provide medical teams with additional information about the severity of a patient's pain, which does not rely on the patient's subjective response.

\subsection{Inter-individual Pain Prediction}
In real-time scenarios, when a new patient visits a hospital, intra-individual level models can not be applied until sufficient data is collected. We report the inter-individual pain prediction results for 47 patients in Table \ref{inter}. The best performance was achieved in Case 4 by DT (accuracy 0.728) compared to 0.429 by MLR in Case 1 by Yang et. al. \cite{yang2018improving}. This indicates that our DT model trained on more vitals signs data collected over a more extended period (five years) from a larger cohort of people could predict the severity of pain for a new patient more accurately. Also, with more data, the models can generalize better, as we did not consider the patient-level differences during both data imputation and pain prediction (Case 4). Additional visit information seemed important in predicting pain scores when considering patient information at data imputation and prediction (Case 1). This indicates that we need to consider the type of visit to predict pain scores from vital signs for a personalized prediction from a generalizable model.

In our original dataset, we have 11 unique self-described pain scores ranging from 0 to 10. It is challenging for one person to distinguish between such broad and granular pain intensity levels and be consistent with every pain episode. 
Hence, we reported our model performance at an inter-individual level by transforming our dataset on a 6-point rating scale, a 4-point rating scale, and a binary rating scale \cite{yang2018improving} in Table \ref{inter_scale}. 
The higher accuracy associated with the narrow scales is attributed to the narrow space to misclassify many records by our models, thereby improving the chances of correctly predicting a pain score. 

We believe the lower performance of RF compared with DT is attributed to the replacement-with-duplicates-based bootstrap approach to sub-sampling used in training a Random Forest model that could lead to training records not representative of the test sample the model is tested on. Furthermore, as a bagging approach to ensemble models, the misclassification error from the first bootstrap sample in random forests is not used to improve a model trained on a different bootstrap sample. Finally, test accuracy is computed by taking the average of the different bootstrap samples' accuracy where bad bootstrap samples might harm the aggregate test accuracy. With DT, however, it is possible that our model captured the ideal training records for a given test set, thereby yielding better accuracy.

\subsection{Pain Change Prediction}
Additional information about the change in pain (increase/decrease/no change) would help determine the effectiveness of therapy and the consideration for management, such as either giving more pain medication, keeping a medication dosage stable or decreasing a pain dosage. Predicting a change in pain may be more critical than having an estimate of the pain score since the medical team can make treatment decisions based on this information and ultimately improve a patient's pain more quickly. Hence, we formulate a three-class pain change classification problem, i.e., increase, decrease, and no change. In this case, a baseline chance accuracy can be 0.33 (1/3). We report the results of our two best performing DT and kNN based inter-individual level classifiers in Table \ref{pain_change}. It is not surprising that our DT model was able to predict a pain change correctly 52.2\% times (0.7\% more) when provided with additional visit information. It indicates our model learned that the change in pain severity might be different for each type of visit for different patients (i.e., Case 1). It is essential to consider whether a visit is an inpatient or outpatient visit to estimate a change in pain intensity for each patient.


\section{Conclusion}
\label{concl}
In this study, we leveraged multiple machine learning algorithms on six physiological measures of patients with SCD to predict pain scores. We were able to deal with missing data and conduct a series of experiments at both intra-individual and inter-individual levels. In each of the experiments, we observed higher accuracy with an increase in data. All the models were able to capture the variation in the type of visits at an inter-individual level when considering the diversity between patients in data imputation and prediction. Our results show Decision Tree as the most promising model, followed by k-Nearest Neighbours and Support Vector Machines. The evaluation demonstrates that using objective physiological measurements to predict subjective pain in SCD patients may be generalizable for a larger cohort of patients. In the future, we look forward to extending our work to visit level analysis and exploring the patients' medication information.


%
\bibliographystyle{splncs04}
\bibliography{sample}

\end{document}